\documentclass[twocolumn,english]{revtex4}
\usepackage[T1]{fontenc}
\usepackage[latin1]{inputenc}
\usepackage{graphicx}
\newcommand{\rr}{{\bf r}}

\newcommand{\vv}{{\bf v}}

\makeatletter

\usepackage{babel}
\makeatother

\begin{document}

\title{Computer Simulation of Cytoskeleton-Induced Blebbing in Lipid Membranes
~\footnote{To appear in  Physical Review E }}

\author{Eric J. Spangler}
\affiliation{Department of Physics, The University of Memphis, Memphis, TN 38152, USA}

\author{Cameron W. Harvey~\footnote{Present address: Department of Physics, University of Notre Dame, Notre Dame, IN 46556, USA}}
\affiliation{Department of Physics, The University of Memphis, Memphis, TN 38152, USA}

\author{Joel  D. Revalee~\footnote{Present address: Department of Physics, University of Michigan, Ann Arbor, MI 48109, USA}}
\affiliation{Department of Physics, The University of Memphis, Memphis, TN 38152, USA}

\author{P.B. Sunil Kumar}
\affiliation{Department of Physics, Indian Institute of Technology Madras, Chennai - 600 036, India
and MEMPHYS - Center for Biomembrane Physics, University of Southern
Denmark, DK-5230, Denmark}

\author{Mohamed Laradji~\footnote{Author to whom correspondence should be addressed. Electronic mail: mlaradji@memphis.edu}}
\affiliation{Department of Physics, The University of Memphis, Memphis, TN 38152, USA
and MEMPHYS - Center for Biomembrane Physics, University of Southern
Denmark, DK-5230, Denmark}
\begin{abstract}
Blebs are balloon-shaped membrane protrusions that form during many physiological processes.  Using computer simulation of a particle-based model for self-assembled lipid bilayers coupled to an elastic meshwork, we investigated the phase behavior and kinetics of  blebbing.  We found that blebs form for large values of   the ratio between the areas of the bilayer and the cytoskeleton. We also found  that blebbing can be induced when the cytoskeleton is subject to a localized ablation or a uniform compression. The results obtained are qualitatively in agreement with the experimental evidence and the model opens up the possibility to study the  kinetics of bleb formation in detail.
\end{abstract}
\maketitle

\newpage{}

\section{Introduction}
It is well established that morphological deformations of the plasma membrane may be induced by the interplay between the cortical cytoskeleton (CSK) and its lipid bilayer (LB) ~\cite{mohandas94,lim02}.  
Many  cells  exhibit transient exoplasmic protrusions, known as blebs,
during several physiological processes, including cytokinesis, cell motility, apoptosis, and virus uptake~\cite{burton97,fackler08,mills98,foller08,jewel82,barros03,mercer08}. 
Recent experiments revealed that blebs can be induced in suspended fibroblasts when these are subjected to a localized ablation of the cortical CSK~\cite{tinevez09}.  Blebbing is also observed when tension on the cortical CSK  is  increased  
after activation of myosin-II motors ~\cite{tinevez09,paluch06}.  
This letter presents a  computational study  of membrane blebbing using a particle-based model for a self-assembled LB coupled to an explicit CSK. 

Blebs are believed to be triggered by a biochemical inhibition of myosin-II leading to either a localized rupture of the cortical CSK or its detachment from bilayer~\cite{cunningham95,charras05,sheetz06,charras08,tinevez09,paluch05}. They then grow into 
spherical protrusions, up to 2 $\mu{\rm m}$ in diameter, which are devoid of acto-myosin meshwork~\cite{sheetz06}. In non-apoptotic cells, blebs retract due to recruitment of actin-membrane linkers to the bleb's membrane,  polymerization and contraction of actin within the bleb~\cite{charras08}. Red boold cells (RBCs) were also shown to exhibit blebbing during their aging. Indeed,  a large fraction of an RBC's plasma membrane is shed into small vesicles that are devoid of spectrin CSK~\cite{backman98,gov05}.  The small size of RBC-shed vesicles suggests that the precursor blebs may have a size about the corral size of the RBC spectrin meshwork~\cite{sheetz06}.  Blebbing in RBCs is believed to result from a reduction of ATP levels during the aging process leading
to a contraction of the sepectrin meshwork~\cite{gov05}.

Despite its importance  to various biological processes, few theoretical studies were performed to understand blebbing~\cite{sens07,tinevez09,young10}.  
A theoretical or computational study of cell blebbing is obviously very difficult due to the vast complexity of cells. Simplified models are  thus needed to infer some aspects of blebbing.  Here,  we propose a particle-based model system, where the only ingredients taken into account are the LB  and a simplified CSK.  Our results clearly  demonstrate that membrane blebbing results from a subtle interplay between contractile tension of the CSK meshwork and the LB elasticity. 

\section{Model and Method}

\begin{figure}[ht]
  \includegraphics[width=2.5in,angle=0]{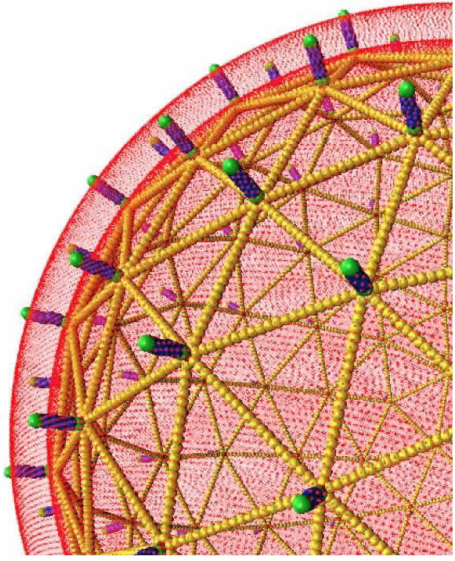} 
    \caption{(Color online) Lipids particles are represented by dots. The cytoskeleton particles are yellow (lightest) beads. The anchoring (bola) lipids have green (middle gray) head particles and blue (darkest gray) hydrophobic beads.}
   \label{fig:model}
\end{figure}

Our study uses a mesoscale implicit-solvent model,  recently developed by us~\cite{revalee08}, of self-assembled lipid molecules with soft interactions.
Here, lipid molecules are coarse-grained into semi-flexible amphiphiles composed of one hydrophilic particle ($h$) connected to a chain of two hydrophobic ($t$) particles. 
The self-assembled one-component lipid vesicles do not possess a spontaneous curvature.
The inner side of the vesicle is underlined with a semi-flexible polymer meshwork tessellated by triangles formed by linking vertices with semi-flexible polymer chains.  The CSK configuration is composed of 162 vertices, corresponding to ${\cal N}_{\rm cor}=320$ triangular corrals. 150 vertices have six  links while 12  have five links. Each link is composed of 8 to 22 monomers. All links are identical for a given vesicle. The vertices are anchored to the membrane through bola lipids mimicking the  anchoring protein complexes in RBCs.  We note that the CSK topology remains conserved during a simulation. A pictorial presentation of the model is shown in Fig.~\ref{fig:model}.

The interaction potential between particles includes two-body interactions, harmonic interactions for the bonds  within a lipid or CSK meshwork, and three-body interactions to account for bending rigidity of  lipid particles and CSK~\cite{revalee08}
\begin{eqnarray}\label{eq:net_potential}
U\left(\left\{{\bf r}_i\right\}\right)&=&\sum_{i,j}U^{\alpha_i\alpha_j}_0\left(r_{ij}\right)+\sum_i U^{\alpha_i}_{\rm bond}\left(r_{i,i+1}\right) \nonumber \\
&+&\sum_i U^{\alpha_i}_{\rm bend}\left({\bf r}_{i-1},{\bf r}_i,{\bf r}_{i+1}\right),
\end{eqnarray}
where ${\bf r}_i$ is the position of particle $i$, $r_{ij}=|{\bf r}_j-{\bf r}_i|$, and $\alpha_i=h$, $t$, or $c$ for a lipid head,  lipid tail, or a CSK bead, respectively.
In Eq.~(\ref{eq:net_potential}),
$U_0^{\alpha_i\alpha_j}$ is a soft two-body interaction between two particles
$i$ and $j$  and is given by
\begin{widetext}
\begin{eqnarray}\label{eq:potential_2body}
U^{\alpha\beta}_0\left(r \right)= 
\left\{\begin{array}{cc}
\left(U^{\alpha\beta}_\mathrm{max}-U^{\alpha\beta}_\mathrm{min}\right)\frac{\left(r_m-r\right)^2}{r_m^2}
+U^{\alpha\beta}_\mathrm{min}  &\mathrm {if\ } r\leq r_m  \\
-2 U^{\alpha\beta}_\mathrm{min}\frac{\left(r_c-r\right)^3}{\left(r_c-r_m\right)^3}+3 U^{\alpha\beta}_\mathrm{min}\frac{\left(r_c-r\right)^2}{\left(r_c-r_m\right)^2} & \mathrm{if}\ \  r_m < r \leq r_c \\
0 &\mathrm{if}\  \  r>r_c, \end{array}\right.
\end{eqnarray}
\end{widetext}
The self-assembly of the lipid chains in this model is achieved through the addition of an attractive interaction between tail particles. We choose $U_\mathrm{min}^{\alpha\beta}=0$
if either $\alpha$ or $\beta=h$, and $U_\mathrm{min}^{\alpha\beta}<0$ if $\alpha=\beta=t$.  $U_\mathrm{max}^{\alpha\beta}>0$ for any values of $\alpha$ or $\beta$. The interaction between the cytoskeleton particles and lipid particles is
assumed repulsive.

The potential $U^\alpha_\mathrm{bond}$ ensures the connectivity between two consecutive monomers in a lipid chain or  cytsoskeleton and is given by
\begin{equation}\label{eq:potential_bond}
U^\alpha_\mathrm{bond}(r)=\frac{k^\alpha_\mathrm{bond}}{2}\left(r-a_b\right)^2,
\end{equation}
where $k^\alpha_\mathrm{bond}$ is the bond stiffness coefficient and $a_b$ is the preferred bond length.  
$U^\alpha_\mathrm{bend}$ is a three-body interaction potential ensuring bending stiffness of the lipid chains and is given by
\begin{equation}\label{eq:potential_bend}
U^\alpha_\mathrm{bend}\left({\bf r}_{i-1},{\bf r}_i,{\bf r}_{i+1}\right)=\frac{k^\alpha_\mathrm{bend}}{2}
\left(\cos\theta_0^\alpha-\frac{{\bf r}_{i,i-1}\cdot{\bf r}_{i,i+1}}{r_{i,i-1}r_{i,i+1}}\right)^2,
\end{equation}
where $k^\alpha_\mathrm{bend}$ is the bending stiffness coefficient of a lipid chain or CSK and $\theta_0^\alpha$ is the preferred splay angle and is chosen as $180^\mathrm{o}$ for all triplets.
The values of the interaction parameters are given by
\begin{eqnarray}
U_{\rm max}^{hh}& = & U_{\rm max}^{cc}=U_{\rm max}^{ht}=U_{\rm max}^{hc}=U_{\rm max}^{tc}=100\epsilon, \nonumber \\
U_{\rm max}^{tt}   & = &200\epsilon, \nonumber \\
U_{\rm min}^{hh} & =& U_{\rm m}^{cc}=U_{\rm min}^{ht}=U_{\rm min}^{hc}=U_{\rm min}^{tc}=0, \nonumber \\
U_{\rm min}^{tt}    & = &-6\epsilon, \nonumber \\
k_{\rm bond}^h    & = & k_{\rm bond}^t=k_{\rm bond}^c=100\epsilon/r_m^2, \nonumber \\
k_{\rm bend}^t      & =  & k_{\rm bend}^c=100\epsilon,
\end{eqnarray}
and $r_c=2 r_m$, and $a_b=0.7r_m$.  

The thickness of the lipid bilayer in the present model is given $4.1 r_m$. The thickness of a in the fluid phase and {\em in vitro} condition is about $4.5 \ {\rm nm}$.  We thus estimate $r_m\simeq 0.91\ {\rm nm}$. The area per lipid for a tensionless bilayer in the present model is about $0.65\ r_m^2$. The area per lipid for a dipalmitoylphosphatidylcholine (DPPC) bilayer in the fluid phase is about $0.63 {\rm nm^2}$~\cite{petrache00}. 
Therefore, we estimate $r_m\simeq 1 {\rm nm}$.
Phospholipid bilayers in the fluid phase and {\em in vitro} typically have a diffusion coefficient
$D \approx 1-10 \times 10^{-12}\ {\rm m^2/s}$~\cite{filippov03}. In our model, $D \approx 0.015 \ r_m^2/\tau$. Therefore, the time scale $\tau\approx 1 - 10 \ {\rm ns}$.  Many of our simulations were performed over durations of milliseconds.
We note that the thickness of the lipid bilayer of the plasma membrane {\em in vivo} is not known, although we do not expect it to be much different from that {\em in vitro}.  We also note that {\em in vivo}, the diffusivity is about one order of magnitude smaller.

The particles are moved using molecular dynamics with a Langevin thermostat~\cite{revalee08}:
$\dot\rr_{i}(t) =  \vv_{i}(t)$ and
$m\dot\vv_{i}(t) = -\nabla_{i}U-\Gamma\vv_{i}(t)+{\bf W}_{i}(t)$,
where $m$ is the mass of a particle (same for all particles).
$\Gamma$ is a bead's friction coefficient,
and ${\bf W}_{i}(t)$ is a random force originating from the heat bath, and satisfies
$\langle{\bf W}_{i}(t)\rangle  =  0$, 
and $\langle{\bf W}_{i}\left(t\right)\cdot{\bf W}_{j}\left(t'\right)\rangle  =  6k_{{\rm B}}T\Gamma\delta_{i,j}\delta\left(t-t'\right)$.
Equations of motion are integrated using the velocity Verlet algorithm with $\Gamma=\sqrt{6} m/\tau$ where the timescale $\tau=r_m(m/\epsilon)^{1/2}$ with $r_m$ and $\epsilon$ being the length and energy scales. Simulations are performed at  $k_{\rm B}T=3\epsilon$ with  $\Delta t =0.02\tau$. 
Many simulations were run  up to  $8\times 10^6$ time steps, which corresponds to  time scales around  $10 {\rm ms}$.
A large number of vesicles were investigated with number of lipids ranging between $35\ 000$ and  $2.5\times 10^5$, corresponding to vesicles' diameter between 70 and 160 nm.
\section{Results}

\begin{figure}[ht] 
  \includegraphics[width=3.8in,angle=0]{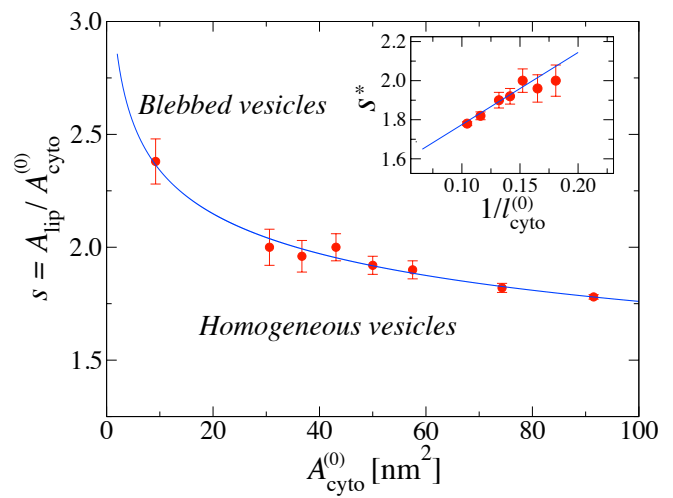} 
    \caption{(Color online) Blebbing phase diagram for closed vesicles in terms of   corral rest area, ${\cal A}_{\rm cyto}^{(0)}$, and  mismatch parameter, $s$.
    The inset shows the transition line, $s^\star$, versus $1/l_{\rm cyto}^{0}$, where  $l_{\rm cyto}^{0}$ is the linear size of a stress-free CSK corral. The solid line is a fit to the equation for $S^{\star}$ given in the text. 
    $A_{\rm cyto}^{(0)}$ is varied by changing the number of beads per link in the cytoskeleton.}
   \label{fig:phase_diagram}
\end{figure}

We first focus on the phase diagram of the system (cf.  Fig.~\ref{fig:phase_diagram}), which is described in terms of the rest area of a CSK corral, ${\cal A}_{\rm cyto}^{(0)}$ (defined as the corral area with minimum elastic energy), and the area mismatch parameter, $s={\cal A}_{\rm lip}/{\cal A}^{(0)}_{\rm cyto}$, where ${\cal A}_{\rm lip}$ is the area of the  LB normalized by the number of corrals, ${\cal N}_{\rm cor}$.
Fig.~\ref{fig:phase_diagram}  indicates a phase transition line, denoted by $s^{\star}({\cal A}_{\rm cyto}^{(0)})$, between a phase where  the vesicle is conformationally homogeneous and a phase
where the vesicle is blebbed. Within the homogeneous phase, the CSK conforms to the vesicle, while in the blebbed phase, the bleb is devoid of the CSK.
To understand this phase behavior and the monotonic decrease  of   $s^{\star}$ with ${\cal A}_{\rm cyto}^{(0)}$ we use an earlier argument by Sens and Gov~\cite{sens07}. In the homogeneous phase, the free energy of the vesicle (assumed to be spherical) is dominated by the curvature energy of the bilayer and the elastic energy of the CSK patches, ${\cal F}_{\rm h}\simeq8\pi\kappa+ (1/2){\cal N}_{\rm cor}e{\cal A}_{\rm cyto}^{(0)}\left ({\cal A}_{\rm cyto}/{\cal A}_{\rm cyto}^{(0)}-1\right)^2$,  where $\kappa$ is the LB bending modulus, $e$ is the corral stretch modulus and ${\cal A}_{\rm cyto}$ is the area of a single stretched corral.
However, if the vesicle exhibits a single bleb, the CSK is stress-free, and the vesicle's free energy is then dominated by its curvature energy,
${\cal F}_{\rm b}\simeq 16\pi\kappa\left(1-{\cal A}_{\rm cyto}^{(0)}/8\pi R_v^2-{\cal A}_{\rm cyto}^{(0)}/8\pi R_b^2\right)$, where $R_v$ and $R_b$ are the radii of the main part of the vesicle (with the subjacent CSK) and the bleb, respectively, both assumed to have almost spherical shape. In the expression of energy of the blebbed vesicle, it is assumed that the area of the bleb's neck is that of a relaxed corral. The transition line, in the case of  a vesicle with a large number of 
corrals (${\cal N}_{\rm cor}\gg1$), obtained by equating the free energies of the homogeneous and the blebbed phases, is then given by $s^\star\approx 1+\sqrt{16\pi\kappa/{\cal N}_{\rm cor}e {\cal A}_{\rm cyto}^{(0)}}$. 
Therefore, the transition line decays monotonically with ${\cal A}_{\rm cyto}^{(0)}$. This
analysis also predicts that  $s^\star\sim 1/l_{\rm cyto}^{(0)}$, where $l_{\rm cyto}^{(0)}$ is the linear size of a stress-free coral. This seems to be in agreement with our results shown in the inset of  Fig.~\ref{fig:phase_diagram},  although
simulations on much larger vesicles is needed to verify whether this linear relation will hold still.

\begin{figure}[ht] 
  \includegraphics[width=3.5in,angle=0]{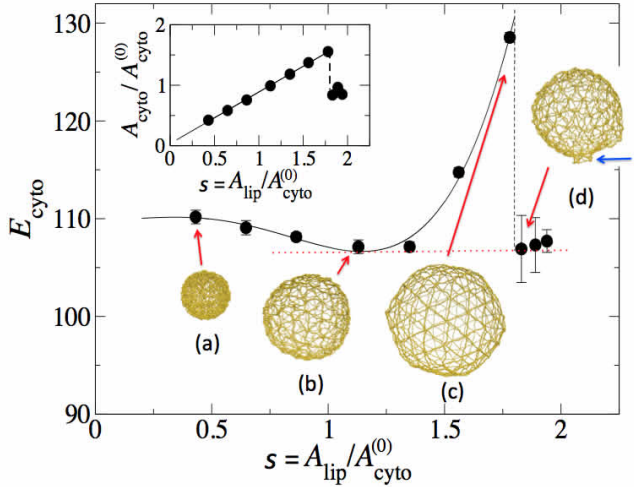} 
    \caption{(Color online) CSK energy, per strand, vs. mismatch parameter, $s$, for the case of ${\cal A}_{\rm cyto}^{(0)}=92 r_{m}^2$ 
     Inset shows the coral area, normalized by its rest area, vs.  $s$, for same system. Blue (most right) arrow next to (d) points to the location of the bleb's neck.}
   \label{fig:elastic_energy_N18}
\end{figure}

To further investigate the nature of blebbing, the CSK elastic energy  is shown {\em vs.} the  area mismatch, $s$, in Fig.~\ref{fig:elastic_energy_N18} for the case of  
${\cal A}_{\rm cyto}^{(0)}=92 r_{m}^2$, together with snapshots of the CSK for different values of $s$. This figure  indicates that the CSK    potential energy  is minimized at $s_{\rm min}\approx 1$ slightly larger than 1. $s_{\rm min}$ 
is slightly above  1 due to the slight curvature of the LB.
When $s< s_{\rm min}$, the CSK is compressed due to the small area of the vesicle, while for $ s_{\rm min}<s<s^\star$, the CSK is stretched, implying that it exerts a  
compressive stress on the LB as $s$ is increased towards the phase transition. In this regime, the increased elastic energy of the CSK is compensated by the low curvature energy of the vesicle which is almost spherical. 
For $s>s^\star$, the CSK elastic energy  becomes too high if it were to conform to the large membrane.  Instead, the free energy is minimized when the vesicle protrudes a bleb that is devoid of CSK. The CSK then conforms to an effective vesicle with smaller area. In this case, the CSK adopts a stress-free configuration (snapshot (d) of Fig.~\ref{fig:elastic_energy_N18}) which is very similar to the case of a non-blebbed vesicle with a stress-free CSK (snapshot (b)
of Fig.~\ref{fig:elastic_energy_N18}). The sharp discontinuity of the CSK elastic energy implies that the transition is first order, in accord with Ref.~\cite{sens07}.

\begin{figure}[h]
  \includegraphics[width=3.5in,angle=0]{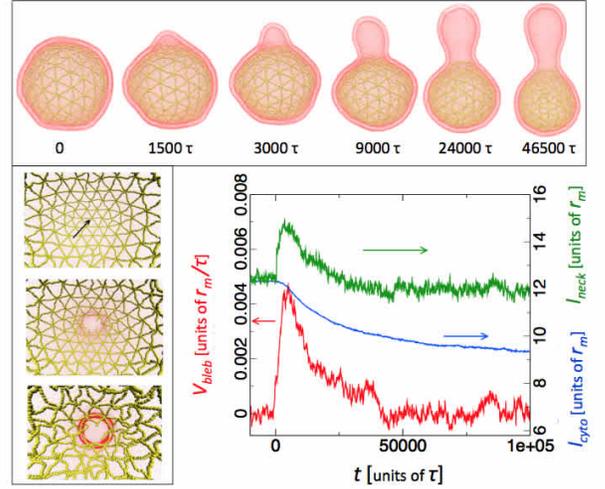} 
    \caption{(Color online) Top panel: Snapshot of a vesicle undergoing blebbing as a result of  a localized ablation of the CSK. The bleb occurs exactly at the location where the damage is made.    Bottom left panel:  Top to bottom snapshots of the vesicle taken from inside, at $t=0$ , $1500$, and $7500\tau$,  respectively. The arrow at $t=0$ indicates the location of the ablated vertex. The dark red (gray) hallow corresponds to the bleb's neck.  Only lipid head particles and CSK particles are shown. 
Bottom right panel:    Bleb front velocity in red (bottom curve), the average length of  a CSK strand between two vertices in blue (middle curve), and the average length of a CSK strand in the bleb's neck region in green (top curve).}
   \label{fig:cut_sequence}
\end{figure}

We now turn to the kinetics of bleb formation and growth following a  localized ablation of the CSK. The numerical experiment shown here was performed on an initially homogeneous vesicle with  corral rest area ${\cal A}_{\rm cyto}^{(0)}=  43.1r_{m}^2$ and mismatch parameter $s=1.90$ (right below the transition line).  At $t=0$,  an arbitrary vertex  is dissociated from its six links, and the kinetics of the system is then monitored over a  long period of time. A  snapshot time series is shown in Fig.~\ref{fig:cut_sequence} (see as well Movie 1 in Supplementary Material), together with a graph depicting the speed of the bleb's apex, $v_{\rm bleb}=d\Delta h/dt$ ($\Delta h$ is the distance between the bleb's apex and its base),  the diameter of the bleb's neck and the average length of a CSK strand.
The snapshot series indicates that the membrane caps immediately after, and at the same location where, the ablation is made. The cap grows rapidly during early times, as shown by the $v_{\rm bleb}$. The speed reaches a maximum, then decays during later times. The qualitative behavior of the apex's speed is similar to that  reported  by Charras {\rm et al}.~\cite{charras05}.  The diameter of the neck evolves non-monotonically with time: The sudden localized dissociation of one vertex from its
six  neighboring strands causes an imbalance of forces on the six neighboring vertices leading to their displacement outward and then an increase of the size of the strands in the ablation region. Meanwhile, this is accompanied by a compression of the overall CSK,  demonstrated by a decrease in the average length of CSK strands.  
After the initial brief expansion of the neck, during which the bleb has a shape of a cap, the neck starts to retract. The onset of retraction of the CSK strands in the neck region coincides with the transition of the bleb's shape to that of a bud. Interestingly, this transition roughly coincides with a slowing down in the growth rate of the bleb's apex. We associate this slowing down to the fact that the net current of lipids flowing into the bleb is proportional to the perimeter of the neck.
Subsequent kinetics is characterized by slower growth of the bleb and a slow relaxation of the CSK's elastic stress. Blebs growth ceases 
once the CSK reaches its stress-free state.  It is noted that no blebbing is observed following ablation if the initial value of $s$ is much below the transition line. We note that the blebs in our study are much smaller than those reported experimentally~\cite{charras05}. A quantitative comparison between the speeds from our study and that from experiment is premature since
small blebs cannot be resolved in experiments. 
\begin{figure}[ht]
  \includegraphics[width=3.5in,angle=0]{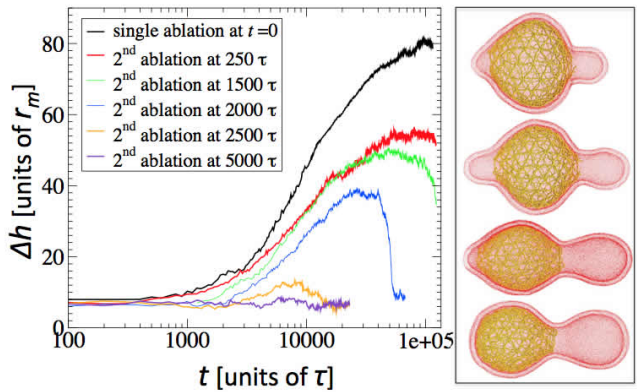} 
    \caption{(Color online) Left panel: Height of the second bleb when a vesicle subjected to two ablations.  Top curve (black) in graph corresponds to the case of a 
a single ablation.  Curves from top to bottom, excluding the first one, correspond to a second ablation at $\Delta t = 250\tau$, $1500\tau$, $2000\tau$, $2500\tau$, and $5000\tau$, respectively. 
Right panel: A snapshot time series of the vesicle for the case of $\Delta t=2000\tau$. Snapshots from top to bottom are taken at $7500$, $15000$, $50000$ and $160000\tau$, respectively.}
       \label{fig:second-ablation}
\end{figure}

We also made a series of computer experiments where a second ablation is subsequently applied at a location diametrically opposite from the first ablation. In Fig.~\ref{fig:second-ablation}, the height of the second bleb, $\Delta h_2$,  is plotted vs. time, together with the height of the growing bleb, $\Delta h_1$, when a single ablation is applied. 
When two ablations are applied simultaneously, then two almost identical blebs form simultaneously and grow at the same rate (data not shown).  
However, when the second ablation is applied subsequently, different kinetics of the second bleb is observed depending on the time lapse, $\Delta t$, between the two ablations. If 
$\Delta t$ is very short ($\lesssim250\tau$), the second bleb forms and grows (cf. red curve in  Fig.~\ref{fig:second-ablation}). However, when $\Delta t\gtrsim 1500\tau$, a second bleb forms and grows during intermediate times (green curve and snapshot series of Fig.~\ref{fig:second-ablation}). During later times, the second bleb then retracts and eventually completely disappears (see Movie 2 in Supplementary Material for the case of $\Delta t = 2000\tau$).  The life time of the second bleb is reduced as the time interval $\Delta t$ is increased.
We note that energetically, one bleb is more favorable than two blebs. Therefore, even for small $\Delta t$, the second bleb should be metastable.
No discernible second bleb is observed when the second ablation is applied after $\Delta t \gtrsim 5000\tau$ (cf. purple line in  Fig.~\ref{fig:second-ablation}). 
These results further substantiate that the observed blebbing  is the result of the interplay between  the CSK elasticity and the lipid bilayer elasticity. We emphasize that in cells, the hydrostatic pressure plays an important role on blebbing, which is not accounted for in the present model. Therefore, the explicit variation of blebbing in our study, in terms of the parameter $s$ is expected to be different from that in cells~\cite{tinevez09}.

\begin{figure}[ht]
  \includegraphics[width=3.5in,angle=0]{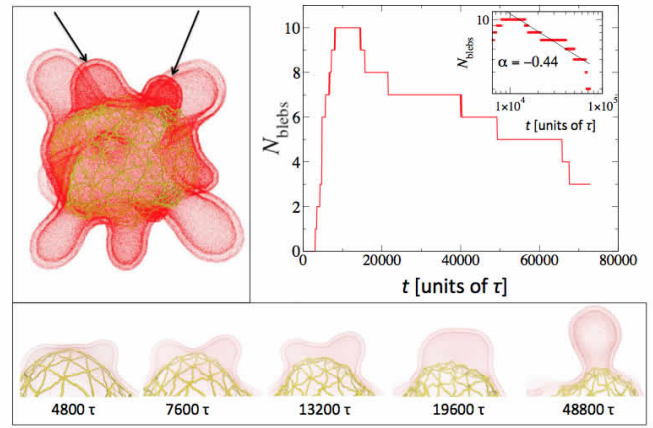} 
    \caption{(Color online) Top left panel: A vesicle, composed of $4\times 10^5$ lipids undergoing blebbing during a uniform contraction of CSK to a point in the phase diagram corresponding to ${\cal A}_{\rm cyto}^0=112 r_m^2$ and $s=2.25$. Top right panel: Number of blebs vs. time. Inset shows same data in a double logarithmic plot. Bottom panel: Sequence illustrating coalescence of two blebs (indicated by arrows on top left panel). }
   \label{fig:coalescence}
\end{figure}

We also looked at a vesicle undergoing a sudden uniform contraction of the CSK, starting from an equilibrium homogeneous state.  This simulates the activity of
myosin motors in  generating a contraction of the actin meshwork. A configuration of the vesicle during blebbing is shown in Fig.~\ref{fig:coalescence}. Right after the sudden CSK's contraction, many blebs are formed throughout the vesicle. 
Fig.~ \ref{fig:coalescence}, indicates a  rapid increase in the number of formed blebs during early times, which then reaches a maximum, and eventually decays at later times. The late time kinetics proceeds through coalescence of blebs and retraction of some blebs, albeit most of blebs coalesce. An example illustrating the coalescence of two neighboring blebs is shown in the snapshot series in Fig.~\ref{fig:coalescence} (also see Movie 3 in Supplementary Material). 
Interestingly, the number of blebs is found to decay as $N_{\rm blebs}\sim t^{\alpha}$ with $\alpha \approx-0.44$ which is  close to $1/2$. Note that a large number of runs and longer simulations are needed to unambiguously extract this exponent.
Since the system is undergoing spinodal decomposition between the blebbed and homogeneous regions. The kinetics of blebs' coalescence is similar to that of a two-component membranes undergoing phase separation without hydrodynamics for which the exponent of the number of buds, $\alpha=-1/2$~\cite{sunil01}. 

\section{Conclusions}
 
In conclusion, we presented a numerical study of blebbing based on a model of a self-assembled LB with explicit CSK. In this model the solvent is accounted for implicitly,  and therefore  volume constraint  and cytosol flow are ignored here. These effects have recently been put forward as essential ingredients of blebbing. 
From investigation of the phase behavior of blebbing, we found that  for small mismatch parameter, $s$ , the equilibrium state is that of a homogeneous vesicle with the CSK conforming to the entire vesicle. However, for relatively large $s$, the equilibrium state is that of a blebbed vesicle, with the bleb being devoid of the CSK. The transition value of $s$ decreases linearly with the linear size of the CSK corral, in agreement with the recent theory of Sens and Gov~\cite{sens07}. 
Blebs are observed when the membrane is subjected to a localized disruption of the CSK or a uniform contraction, in line with experimental observations. 

\section*{Acknowledgements}
ML acknowledges support from the National Science Foundation (DMR-0812470 and DMR-0755447) and the Research Corporation (CC66879).
We thank Prof. Pierre Sens  for stimulating discussions.

\end{document}